\documentclass[aps,prl,reprint,superscriptaddress,floatfix,longbibliography,footinbib]{revtex4-1}
\usepackage{mathtools}
\usepackage{amsfonts}
\usepackage{amsmath,comment}
\usepackage{tabularx}
\usepackage{float}

\usepackage{xcolor}
\newcommand{\be}{\begin{equation}}
\newcommand{\ee}{\end{equation}}
\newcommand{\bea}{\begin{eqnarray}}
\newcommand{\eea}{\end{eqnarray}}

\begin{document}

\title{Rapidity and momentum distributions of 1D dipolar quantum gases}

\author{Kuan-Yu Li}
\altaffiliation[K.-Y.~Li and Y.~Z.~contributed equally to this work.]{}
\affiliation{Department of Applied Physics, Stanford University, Stanford, CA 94305, USA}
\affiliation{E.~L.~Ginzton Laboratory, Stanford University, Stanford, CA 94305, USA}
\author{Yicheng Zhang}
\altaffiliation[K.-Y.~Li and Y.~Z.~contributed equally to this work.]{}
\affiliation{Department of Physics, Pennsylvania State University, University Park, PA 16802, USA}
\affiliation{Homer L. Dodge Department of Physics and Astronomy, The University of Oklahoma, Norman, OK 73019, USA}
\affiliation{Center for Quantum Research and Technology, The University of Oklahoma, Norman, OK 73019, USA}
\author{Kangning Yang}
\affiliation{E.~L.~Ginzton Laboratory, Stanford University, Stanford, CA 94305, USA}
\affiliation{Department of Physics, Stanford University, Stanford, CA 94305, USA}
\author{Kuan-Yu Lin}
\affiliation{E.~L.~Ginzton Laboratory, Stanford University, Stanford, CA 94305, USA}
\affiliation{Department of Physics, Stanford University, Stanford, CA 94305, USA}
\author{\\Sarang Gopalakrishnan}
\affiliation{Department of Physics, Pennsylvania State University, University Park, PA 16802, USA}
\affiliation{Department of Electrical and Computer Engineering, Princeton University, Princeton, NJ 08544, USA}
\author{Marcos Rigol}
\affiliation{Department of Physics, Pennsylvania State University, University Park, PA 16802, USA}
\author{Benjamin L.~Lev}
\affiliation{Department of Applied Physics, Stanford University, Stanford, CA 94305, USA}
\affiliation{E.~L.~Ginzton Laboratory, Stanford University, Stanford, CA 94305, USA}
\affiliation{Department of Physics, Stanford University, Stanford, CA 94305, USA}

\date{\today}

\begin{abstract}

We explore the effect of tunable integrability breaking dipole-dipole interactions in the equilibrium states of highly magnetic 1D Bose gases of dysprosium at low temperatures. We experimentally observe that in the strongly correlated Tonks-Girardeau regime, rapidity and momentum distributions are nearly unaffected by the dipolar interactions.  By contrast, we also observe that significant changes of these distributions occur when decreasing the strength of the contact interactions.  We show that the main experimental observations are captured by modeling the system as an array of 1D gases with only contact interactions, dressed by the contribution of the short-range part of the dipolar interactions. Improvements to theory-experiment correspondence will require new tools tailored to near-integrable models possessing both short and long-range interactions.

\end{abstract}

\maketitle

One-dimensional (1D) bosonic gases with only contact interactions are \emph{integrable}, and, consequently, they possess stable quasiparticles~\cite{Cazalilla2011odb}. Integrability is in general unstable to the addition of long-range interactions. Even weak integrability-breaking interactions have drastic effects on the nonequilibrium dynamics of integrable systems, causing relaxation to a thermal distribution: For dipole-dipole interactions (DDI) these dynamical effects were recently explored~\cite{Tang2018tni}. By contrast, the effects of integrability-breaking interactions on equilibrium states are less clear: Instead of causing quasiparticles to decay, one might expect (in the spirit of Fermi liquid theory) that interactions simply perturbatively dress the quasiparticles. When the energy scale associated with integrability breaking is a small fraction of the other natural energy scales, it is plausible that the dressing will be weak and the bare quasiparticles can still provide an accurate description. This expectation has not been experimentally tested so far; it is not a priori clear how the dressing depends on the parameters of the integrable system and on the type of integrability breaking interaction.

\begin{figure}[!t]
    \centering
    \includegraphics[width=0.95\columnwidth]{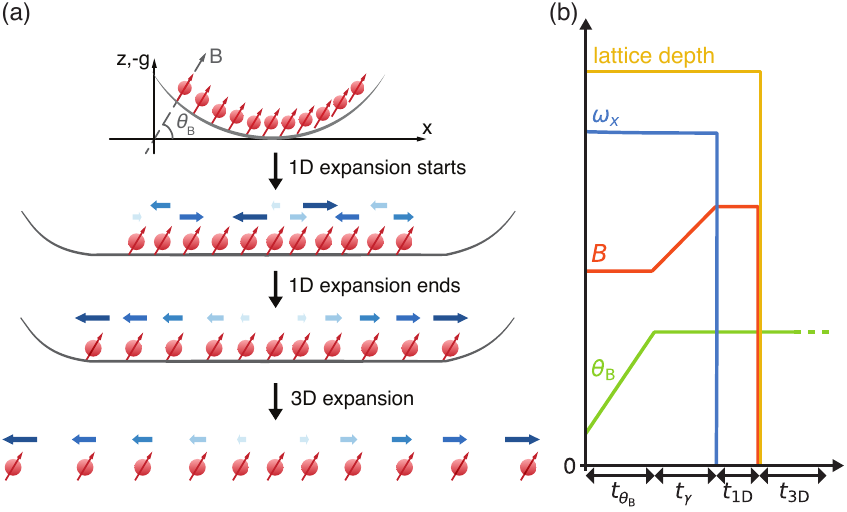}
    \vspace{-0.2cm}
    \caption{(a) Schematic illustrating the experimental sequence for measuring the rapidity distribution of a dipolar 1D gas. A dipolar 1D gas is prepared with a magnetic field magnitude $B$ and angle $\theta_\text{B}$ resulting in contact strength $g_\text{1D}$. Then, the underlying harmonic trap is suddenly removed, while the transverse confinement is maintained. This allows the quantum gas to expand in a flat, 1D trap along $\hat{x}$. Time-of-flight absorption imaging follows 3D expansion by switching off all optical traps. The blue arrows denote the rapidities. (b) Timing sequence for creating a dipolar 1D gas at dipolar angle $\theta_\text{B}$ and $g_\text{1D}$. Once the quantum gas is loaded into a quasi-1D trap, the $B$-field angle is slowly rotated from $55^\circ$ to $\theta_\text{B}=0^\circ, 35^\circ,$ or $90^\circ$ in a time $t_{\theta_\text{B}}$, or kept at $55^\circ$, as the experiment requires. $g_\text{1D}$ is then set to its final value by ramping the $B$-field strength near a Feshbach resonance in a time $t_\gamma = 50$~ms.}
    \label{fig1}
\end{figure} 

In the past, characterizing the dressing of quasiparticles would have been a forbidding experimental challenge: In dense, strongly interacting systems, the mapping between quasiparticles and microscopic particles is nontrivial, and the quantum numbers of the quasiparticles (known as rapidities) are distinct from the microscopic particle momenta, making their distribution hard to measure~\cite{calabrese_essler_review_16, vidmar16}. In a recent experimental breakthrough, a modified time-of-flight (TOF) sequence was developed to measure the rapidity distribution of 1D gases~\cite{wilson_malvania_20, malvania_zhang_21}. In this protocol, one first allows the system to freely expand in 1D under near-integrable dynamics; this step preserves the rapidity distribution. Once the system is dilute, rapidity and momentum distributions coincide, and one can extract the rapidity distribution via TOF imaging.

{Here, we use measurements of rapidity and momentum distributions in an array of 1D bosonic gases with a tunable DDI to explore how the DDI affects the equilibrium properties, e.g., via a dressing of the quasiparticles, and how the effect of the DDI varies when changing the strength of the contact interactions. We find that both distributions are nearly unaffected by the DDI in the Tonks-Girardeau (TG) regime, suggesting that the bare quasiparticles can be used to characterize that regime. As the strength of the contact interactions is decreased, the densities of the 1D gases increase and with them the strength of the dipolar interactions. We find that, as a result, the dressing of the quasiparticles becomes significant and needs to be taken into account in any model of the system in that regime. To attempt to do so, we confirm that modeling the system as an array of 1D gases with only (integrable) contact interactions (dressed with the short-range part of the DDI) is most accurate in the TG regime.  We also show that the model captures the experimental trends as the strength of the contact interactions is decreased. A more accurate correspondence will require the development of new theoretical tools to account for the long-range part of the dipolar interaction and dynamical effects during the initial state preparation.} 

A dipolar $^{162}$Dy BEC of $2.3(1)\times 10^4$ atoms is prepared in a 1064-nm crossed optical dipole trap (ODT) with a final trap frequency of $2\pi\times$[55.5(6), 22.5(5), 119.0(7)]~Hz; more details can be found in Refs.~\cite{Kao2021tpo, Tang2018tni}. The gas is then adiabatically loaded into a 2D optical lattice that is blue-detuned from Dy's 741-nm transition~\cite{Supp, Kao2017ado}. Roughly two thousand 1D traps are populated, with the central ones containing about 40 atoms. Before loading, the angle of an externally imposed magnetic field is set to $\theta_\text{B} \approx 55^\circ$ with respect to $\hat{x}$, the 1D axis of the gases. This minimizes the intratube DDI among atoms within the same 1D trap, which scales as $1-3\cos^2{\theta_\text{B}}$. When the optical lattice reaches a depth of $30 E_\text{R}$, the 1D trap frequency $\omega_x$ is lowered to $2\pi\times 36.4(3)$~Hz by reducing the power of the 1064-nm ODT. The recoil energy is $E_\text{R}=\hbar^2 k_\text{R}^2/2m$ and $k_\text{R}=2\pi/741$~nm, where $m$ is the mass of the highly magnetic, 10 Bohr magneton $^{162}$Dy atom we use~\cite{Chomaz2022dpa}. Simultaneously, a 1560-nm ODT is superimposed to negate the antitrapping potential caused by the blue-detuned optical lattice. We then set $\theta_\text{B}$ to the particular value we require by rotating the magnetic field while maintaining the same lattice depth by adjusting the optical lattice power to compensate for the large tensor light shift~\cite{Kao2017ado}. The 1D-regularized contact interaction strength $g_\text{1D}$ is then ramped to the required final value via a confinement-induced resonance~\cite{Olshanii1998asi, Haller2010cri,Kao2021tpo}.  The resonance is accessed by adjusting the magnitude $B$ of the magnetic field near a Feshbach resonance~\cite{Supp,Baumann2014ool}. To create the dilute TG gases, we prepare a smaller BEC of $5.8(2)\times 10^3$ atoms and set $\theta_\text{B} = 90^\circ$; this leads to a maximum of about 15 atoms in the central 1D tubes.

\begin{figure}[!t]
    \centering
    \includegraphics[width=0.85\columnwidth]{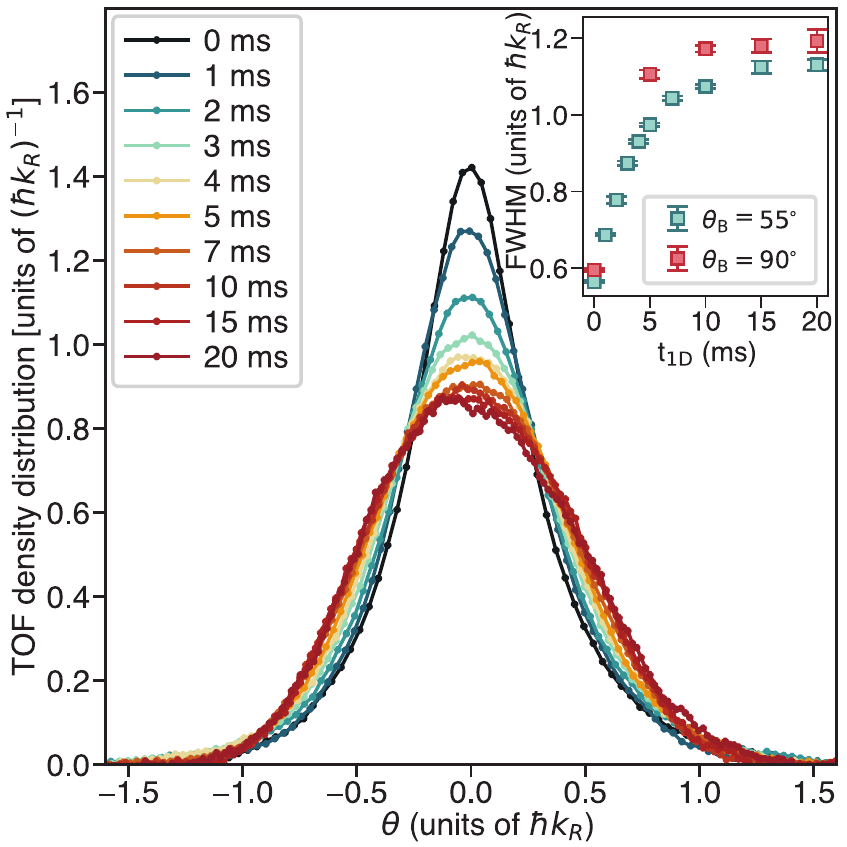}
    \vspace{-0.2cm}
    \caption{The TOF density distribution for $\theta_\text{B}=55^\circ$ and $\gamma_\text{T}\approx 16$ for $t_\text{1D}= 0$ to 20~ms. The width of the distributions' $\theta$ has been scaled by $\hbar k_\text{R}$. 
    Data at times ${>}15$~ms suffer from imaging artifacts and are not used. Inset: The evolution of the FWHM of the distribution versus $t_\text{1D}$ for $\gamma_\text{T}\approx 16$ at $\theta_\text{B}=55^\circ$ (light green) and $\gamma_\text{T}\approx 19$ at $\theta_\text{B}=90^\circ$ (red). Error bars are explained in Ref.~\cite{Supp}.}
    \label{fig2}
\end{figure}

Figure~\ref{fig1} illustrates the experimental sequence for measuring rapidity and momentum distributions. The rapidity distribution is measured using a 1D expansion of duration $t_\text{1D}=15$~ms followed by a 3D expansion of duration $t_\text{3D}=18$~ms. Momentum distributions are measured by setting $t_\text{1D}=0$ with the same $t_\text{3D}$. The magnetic field is switched to the imaging axis $\hat{y}$ at a time 5~ms after the start of the 3D expansion. To reduce the effect of initial gas size, we employ momentum focusing for measuring momentum distributions~\cite{Supp}. To determine the appropriate $t_\text{1D}$ for the rapidity measurement, TOF density distributions with $t_\text{1D}=0-20$~ms are measured; see Fig.~\ref{fig2}. By $t_\text{1D}\approx15$~ms, the distributions have asymptoted to the same shape, indicating that the density distribution reflects the rapidity distribution~\cite{Sutherland_98, rigol_05, wilson_malvania_20}. The inset also shows the saturation in the full width at half maximum (FWHM) of the distribution beyond 10--15~ms. For longer $t_\text{1D}$, imaging artifacts, stray magnetic field gradients, and lower signal-to-noise ratio degrade the image quality, as can be seen in the 20-ms data.  We therefore use $t_\text{1D}=15$~ms for the rapidity measurements.

{Each 1D gas can be described by the Lieb-Liniger Hamiltonian~\cite{lieb_liniger_63} with the addition of an intratube DDI $U^{\rm 1D}_{\rm DDI}$ and a harmonic confining potential $U_{\rm H}$:
\begin{eqnarray}\label{eq:H_LL}
H&=&\sum_{i=1}^{N}\left[-\frac{\hbar^2}{2m}\frac{\partial^2}{\partial x_i^2}+U_{\rm H}(x_i)\right]\\&&+\sum_{1\leq i<j \leq N}\left[g_{\rm 1D}^{\rm vdW}\delta(x_i-x_j)+U^{\rm 1D}_{\rm DDI}(\theta_\text{B}, x_i-x_j)\right]\,,\nonumber
\end{eqnarray}
where $m$ is the atomic mass, $N$ is the number of atoms, and $g_{\rm 1D}^{\rm vdW}$ is the effective 1D contact interaction due to the van der Waals force; see Ref.~\cite{Supp}. Solving this Hamiltonian is very challenging because of the presence of the DDI term. To make theoretical progress, we account for only the leading-order, short-range effect of the intratube DDI. (The intertube DDI is neglected.) Hence, we solve this 1D Hamiltonian after replacing $g_{\rm 1D}^{\rm vdW}\rightarrow g_{\rm 1D}=g_{\rm 1D}^{\rm vdW}+g_{\rm 1D}^{\rm DDI}$ and setting $U^{\rm 1D}_{\rm DDI}(\theta_\text{B}, x_i-x_j)=0$~\cite{Supp}. We note that the properties of the Lieb-Liniger model are parameterized by $\gamma=mg_\text{1D}/n_{\rm 1D}\hbar^2$, where $n_{\rm 1D}$ is the 1D particle density. A $\gamma =1$ denotes a strongly correlated Bose gas of intermediate-strength interactions, while $\gamma \rightarrow \infty$ indicates a TG gas, which can be mapped onto a system of noninteracting spinless fermions~\cite{Cazalilla2011odb}.}

To model the experimental state preparation as closely as possible, we assume there is a lattice depth $U^*_{\rm 2D}$ at which the 3D gas decouples into individual 1D gases as the 2D optical lattice is turned on~\cite{malvania_zhang_21}. {At this lattice depth, we also assume that the 1D gases are in thermal equilibrium with each other at temperature $T^*$ and at a global chemical potential that is set by the total number of particles}. Then, using the local-density approximation (LDA) and the thermodynamic Bethe ansatz (TBA)~\cite{yang_yang_69}, we determine $N$ and the entropy of each 1D gas as functions of $U^*_{\rm 2D}$ and $T^*$. As the depth of the 2D optical lattice is increased beyond $U^*_{\rm 2D}$, we assume that the 1D gases neither exchange particles nor interact.  Thus, they no longer are in thermal equilibrium with each other. We assume this part of the loading process is adiabatic, i.e., that the entropy of each 1D gas is constant. We find the temperature of each 1D gas using: (i) the number of atoms and entropies calculated at decoupling, and (ii) the experimental parameters measured at the end of the state preparation.  The momentum and rapidity distributions are computed using these temperatures. 

We compute the momentum distributions in the presence of the trap using path integral quantum Monte Carlo with worm updates~\cite{boninsegni_prokofev_06a, boninsegni_prokofev_06b, xu_rigol_15}. The rapidity distributions are computed within the LDA by solving the TBA equations. We then sum the results of the 1D gases to compare to the experimental absorption-imaging measurements (which provide distributions averaged over all 1D gases). The values of $\gamma_\text{T}$ reported in the figures and throughout the text reflect the $\gamma$ at which the LL model---at the experimentally set $g_{\rm 1D}$ at {\it finite temperature}---exhibits the same ratio of kinetic--to--interaction energy obtained in our model at $\theta_\text{B}=55^\circ$~\cite{Supp}. 

\begin{figure}[!t]
    \centering
    \includegraphics[width=0.98\columnwidth]{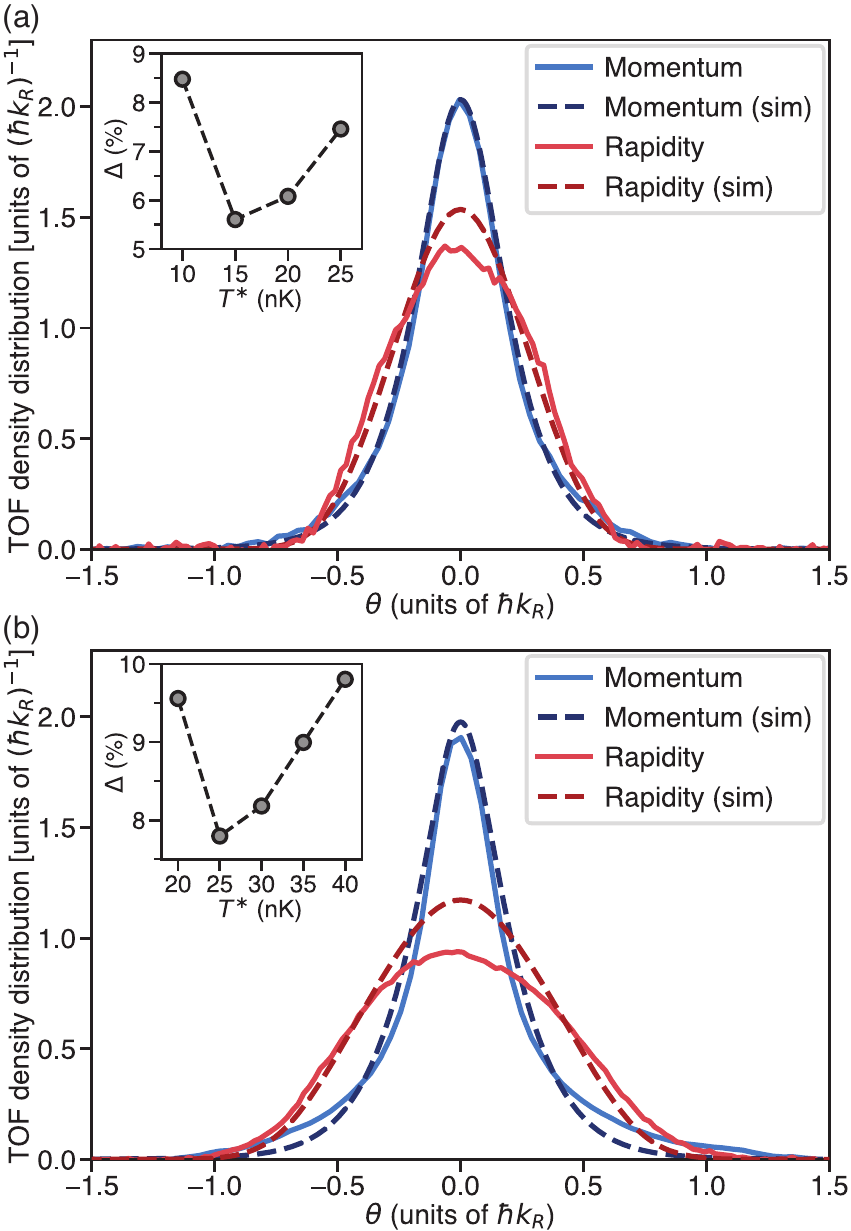}
    \caption{Momentum and rapidity distributions; $\theta$ denotes either momentum or rapidity. Solid lines show the experimental momentum (blue) and rapidity (red) distributions, while the dashed lines show the simulation results. (a) Distributions for $\theta_\text{B}=90^\circ$ and $\gamma_\text{T}\approx 420$ in the TG limit. The simulations use $T^*=15$~nK and $U_\text{2D}^*=5E_\text{R}$. (b) Distributions for $\theta_\text{B}=55^\circ$ and $\gamma_\text{T}\approx 6.7$. Simulations use $T^*=25$~nK and $U_\text{2D}^*=5E_\text{R}$. Insets show the theoretical error used to select $T^*$~\cite{Supp}.}
    \label{fig3}
\end{figure}

\begin{figure*}[!t]
    \centering
    \includegraphics[width=0.9\textwidth]{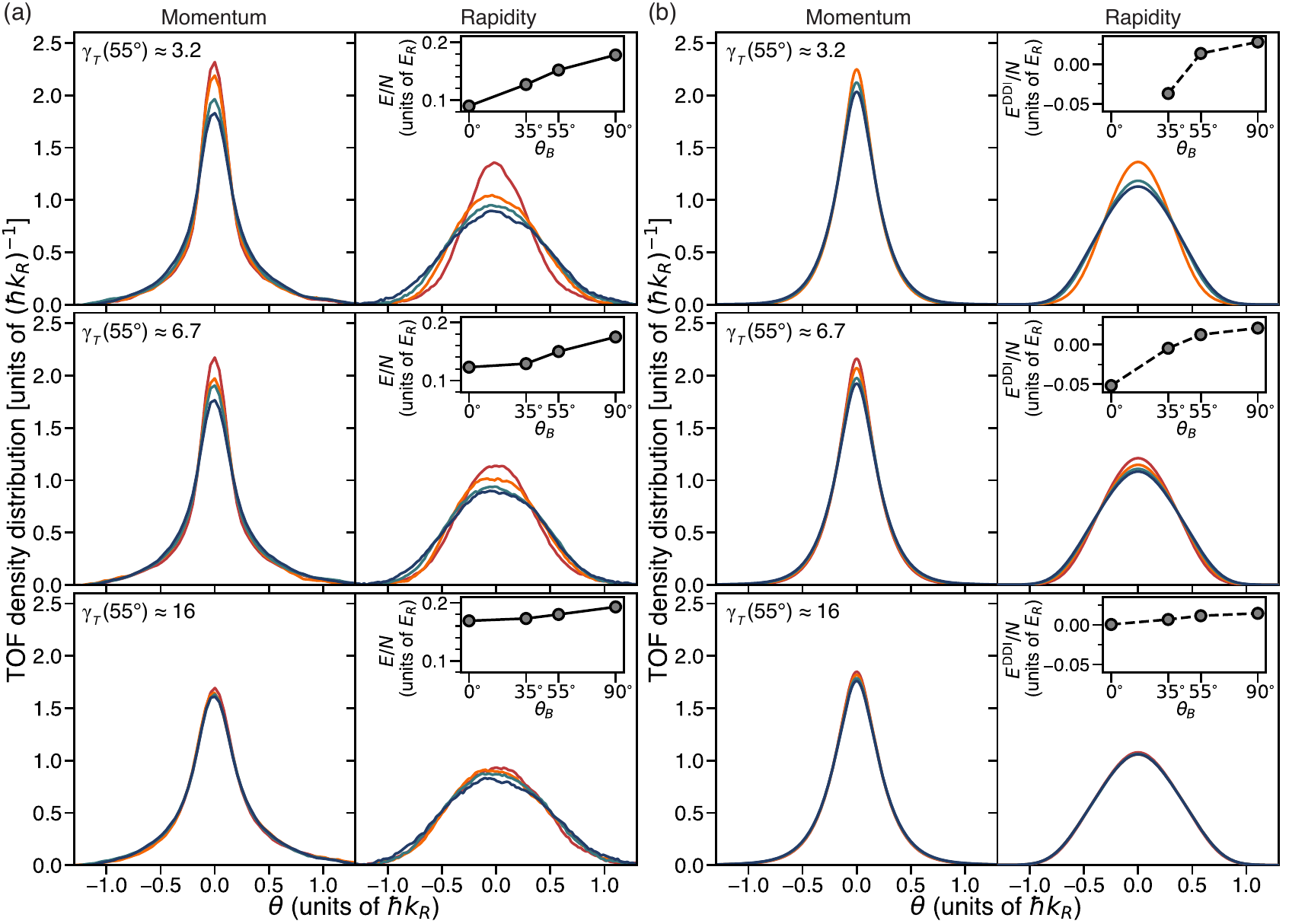}
    \caption{(a) Measured momentum and rapidity distributions at field angles $\theta_\text{B} = 0^\circ$ (red), $35^\circ$ (orange), $55^\circ$ (green), and $90^\circ$ (blue) with $\gamma_\text{T}(\theta_B=55^\circ)\approx 3.2$, 6.7, and 16. $\theta$ denotes either momentum or rapidity. (b) Corresponding simulation curves. {The insets in (a) show the total (interaction plus kinetic) energy that has been experimentally estimated from the rapidity distributions. Insets in (b) show the theoretically estimated total DDI energies, intertube plus short-range intratube.} The $\theta_\text{B}$-dependence comes from the contribution to $g_\text{1D}$ from the short-range part of the intratube 1D DDI. Note that the theory curves are missing for the case of $\gamma_\text{T}(55^\circ)\approx 3.2$ and $\theta_\text{B}=0^\circ$ because $g_\text{1D}$ becomes negative and we cannot simulate that regime. }
    \label{fig4}
\end{figure*}

The free parameters in our model of state preparation are $U^*_{\rm 2D}$ and $T^*$. We find the results to be rather insensitive to the precise value of $U^*_{\rm 2D}$, which suggests that assuming a single decoupling depth for all 1D gases is a reasonable approximation. We select $U^*_{\rm 2D}=5E_\text{R}$~\cite{Supp}. To find $T^*$, we minimize the quadrature sum of the differences between the experimental and theoretical momentum and rapidity distributions~\cite{Supp}, which we call the ``theoretical error.'' We plot this error for the results in the TG limit versus $T^*$ in the inset of Fig.~\ref{fig3}(a). We find the error minimum to be ${\sim}5.5\%$ at $T^*\approx 15$~nK. For $U^*_{\rm 2D}=5E_\text{R}$ and $T^*\approx 15$~nK, we estimate the intertube DDI energy to be ${\sim} 4\%$ of the kinetic + interaction plus trap energy of the 1D gases; see~\cite{Supp} for how intertube DDI energy is calculated. The comparison between the theoretical and experimental results for the momentum and rapidity distributions is shown in Fig.~\ref{fig3}(a). The agreement is remarkable for the momentum distribution. The theoretical rapidity distribution is slightly narrower, which might be due in part to intertube DDI energy being converted into rapidity energy during the 1D expansion.

Similarly, the comparison between experimental and theoretical results for the case of $\gamma_\text{T}\approx 6.7$ and $\theta_\text{B}=55^\circ$ is shown in Fig.~\ref{fig3}(b). For these parameters, we estimate the intertube DDI energy to be ${\sim} 5.5\%$ of the sum of kinetic, interaction, and trap energies of the 1D gases. The theoretical distributions follow the experimental ones, though less closely than in Fig.~\ref{fig3}(a)]; they predict a lower occupation of high momenta and a narrower rapidity distribution. 

Figure~\ref{fig4} reports our main results, the momentum and rapidity distributions of 1D dipolar quantum gases upon changing the contact and DDI strengths (and DDI sign). To reduce systematic variation, we always start from the same state---the one in Fig.~\ref{fig3}(b)---when producing 1D gases with different interactions. That is, we begin with intratube nondipolar gases ($\theta_\text{B}=55^\circ$) at the background scattering length (yielding $\gamma_\text{T}\approx 6.7$) before we then change the magnetic field strength and $\theta_\text{B}$ to the desired final setting. No additional fitting is used to produce the theory curves because the number of atoms and entropies of the 1D gases at decoupling were already computed. Hence, we need to calculate only the temperature of the 1D gases for the experimental parameters after the change in magnetic field angle and/or strength.

{The experimental [theoretical] results are shown in Fig.~\ref{fig4}(a) [Fig.~\ref{fig4}(b)]. The experimentally observed broadening of the momentum and rapidity distributions for increasing $\gamma_\text{T}$ at fixed $\theta_\text{B}$ and/or increasing $\theta_\text{B}$ at fixed $\gamma_\text{T}$ is qualitatively captured by our theoretical model. It can be understood to be the result of increasing total (interaction plus kinetic) and kinetic energies through the increase of $g_{\rm 1D}$ by way of the Feshbach resonance or short-range DDI.} 

{We find that the rapidity and momentum distributions depend weakly on $\theta_\text{B}$ in the strongly correlated ($\gamma_\text{T}=16$) TG regime. They exhibit larger changes versus $\theta_\text{B}$ as $\gamma_\text{T}$ decreases. The insets in Fig.~\ref{fig4}(a) show the changes with $\theta_\text{B}$ of the experimental estimation of the sum of the interaction and kinetic energies, calculated using the measured rapidity distributions. The insets in Fig.~\ref{fig4}(b) show the changes with $\theta_\text{B}$ of our theoretical estimation of the total DDI energies, intertube plus short-range intratube. One can see that the changes in the total energy in the experiment become larger as $\gamma_\text{T}$ decreases, and they parallel the larger changes observed in the estimated DDI energy. Our results illustrate how the nature of the equilibrium state changes as one tunes $\gamma_\text{T}$. When the contact interactions are strong, particles avoid each other and the 1D densities are lower, resulting in weaker dipolar interactions. As one decreases $\gamma_\text{T}$, particles in the equilibrium state of the integrable system are likelier to overlap; therefore, the 1D densities increase, and with them, the strength of the DDI. Remarkably, all these variations are accessible in our experimental apparatus.}

In summary, we showed that the DDI significantly effects the equilibrium rapidity and momentum distributions of dipolar $^{162}$Dy gases as one departs from the strongly correlated TG regime{, suggesting that an increasingly stronger dressing of the quasiparticles takes place}. Our model captured the main experimental trends, but quantitative differences remain. This is likely due, in part, to not accounting for the effect of the long-range aspect of the DDI. It couples different 1D gases as well as bosons that are far away within each 1D gas. The long-range DDI produce  correlations and slow dynamical processes that go beyond what can be computed using state-of-the-art numerical methods. Another potential source of discrepancy are the nonthermal effects related to the near-integrability of the 1D gases as well as to heating, which we neglected. {Nevertheless, it is remarkable that we are able to closely describe the experimental results in such a complex, strongly-interacting system despite the above-mentioned omissions in our modeling.  Notably, our largest ``theoretical error'' for the results reported in Fig.~\ref{fig4} is only ${\sim}11\%$.}

We hope our findings will motivate studies to incorporate the long-range part of the DDI in arrays of 1D gases described by otherwise integrable models~\cite{Panfil2023toi}, {both to understand and quantify how they dress the quasiparticles in equilibrium} and to clarify the role of near-integrability nonthermal effects during the initial state preparation. Such endeavors may usher a new direction for precision quantum many-body physics involving near-integrable models with short-range interactions perturbed by long-range interactions.

\begin{acknowledgments}
We thank Wil Kao for early experimental assistance and acknowledge the NSF (PHY-2006149) and AFOSR (FA9550-22-1-0366) for funding support. K.-Y.~Lin acknowledges partial support from the Olympiad Scholarship from the Taiwan Ministry of Education. Y.Z.~and M.R.~acknowledge support from the NSF Grant No.~PHY-2012145. S.G.~acknowledges funding from the James S.~McDonnell and Simons Foundations and an NSF Career Award. Y.Z.~acknowledges support from Dodge Family Postdoctoral Research Fellowship at the University of Oklahoma.
\end{acknowledgments}

\bibliography{final}

\vspace{1cm}
\newpage
\setcounter{figure}{0}
\setcounter{equation}{0}
\setcounter{section}{0}

\renewcommand{\thetable}{S\arabic{table}}
\renewcommand{\thefigure}{S\arabic{figure}}
\renewcommand{\theequation}{S\arabic{equation}}
\renewcommand{\thesection}{S\arabic{section}}

\onecolumngrid

\begin{center}

{\large \bf Supplemental Material:}\\

\vspace{0.3cm}

\end{center}

\twocolumngrid

\label{pagesupp}

\setcounter{figure}{0}  

\renewcommand{\figurename}{Supp.~Fig.}

\section{Experiments}

\subsection{Experiment sequence}

\subsubsection{BEC production}

A $^{162}$Dy dipolar Bose-Einstein condensate (BEC) is prepared by evaporative cooling in a 1064-nm crossed optical dipole trap (ODT) with beam waists ${\sim}65$~$\mu$m along $\hat{x}$. The typical atom number is $2.3(1)\times 10^4$ at a temperature of 38~nK with a BEC fraction of near 75\%; these numbers are consistent with those reported in Ref.~\cite{Tang2015bco}. The final ODT trap frequency is [55.5(6), 22.5(5), 119.0(7)]~Hz. The bias magnetic field is set along $\hat{z}$ during the evaporation; this is the $\theta_\text{B} = 90^\circ$ direction, where $\theta_\text{B}$ is the field angle with respect to $\hat{x}$.  The field is then slowly rotated to $\theta_\text{B}=55^\circ$. During the field rotation, the field vector is kept in the $x$-$z$ plane with a constant field magnitude. The angle $\theta_\text{B}=55^\circ$ is chosen to remove the intratube dipole-dipole interaction (DDI) so that the results can be compared with numerical simulations using the Lieb-Liniger model. For the Tonks-Girardeau (TG) results, we prepare a BEC of $5.8(2)\times 10^3$ atoms at a temperature of ${\sim}20$~nK with the field angle fixed at $90^\circ$ for better field stability as we approach the Feshbach resonance~\cite{Kao2021tpo}.

\subsubsection{Lattice loading}

The BEC is loaded into a 2D optical lattice to create an array of 1D gases while the 1064-nm ODT remains on. The 2D optical lattice is 5-GHz blue-detuned from the $\lambda=741$~nm atomic transition. The 741-nm beams have a beam of waist ${\sim}$150~$\mu$m and are retroreflected. The lattice depth is adiabatically ramped to 30$E_\text{R}$ in 200~ms, creating a strong confinement in both $\hat{y}$ and $\hat{z}$. The resulting transverse trap frequency is $\omega_\perp=2\pi\times 25$~kHz. $E_\text{R}=\hbar^2 k_R^2/2m$ is the recoil energy and $k_R=2\pi/\lambda$.   There are around 40 atoms in the center tube.  This reduces to around 15 for the TG case.

\subsubsection{1D harmonic and flat trap}

A 1560-nm ODT is then superimposed onto the 1064-nm ODT. Their powers are adjusted such that the longitudinal trap frequency is $\omega_{||}=2\pi\times 36.4(3)$~Hz. Using two ODT beams allows us to have better control of the trap shape when switching from the harmonic trap to the flat trap.

The longitudinal antitrap frequency due to the blue-detuned lattice beams is $\sim$7~Hz at a lattice depth of 30$E_\text{R}$. This antitrapping potential is balanced by the 1560-nm ODT. Its beam waist is 150~$\mu$m to match the shape of optical lattice beams. This results in a flat, 1D trap of length 60~$\mu$m. The final trap configuration consists of the superposition of the 1064-nm ODT, the blue-detuned, 741-nm 2D optical lattice, and the 1560-nm ODT. The more tightly confining 1064-nm ODT sets the final longitudinal trap frequency $\omega_{||}$. This configuration allows us to quickly switch off the longitudinal harmonic trap by turning off the 1064-nm ODT.

\subsubsection{Field rotation and magnitude ramping}

After the array of 1D dipolar gases is created, the field is rotated from $\theta_\text{B}=55^\circ$ to the final angle of choice, either $\theta_\text{B}=0^\circ,\,35^\circ$ or $90^\circ$, or kept constant at $55^\circ$. The magnetic field is held within the $x$-$z$ plane throughout the rotation at the evaporation field magnitude. The angular velocity is chosen to be $4\pi/3$ rad/s, which we experimentally found to be sufficiently slow to avoid collective excitations. Because of strong tensor light shifts~\cite{Kao2017ado}, we must adjust the lattice power during the field rotation to maintain the 30$E_\text{R}$ lattice depth. After the field reaches the final angle, we ramp its magnitude to set $\gamma$. This takes 50~ms. For the TG case, we hold the field at $\theta_\text{B}=90^\circ$ throughout the sequence. The final field ramp takes 30 ms to reach the desired $\gamma$.  This value is chosen to avoid atom loss during the ramp and to minimize collective excitations.

\subsubsection{Scattering length and Feshbach resonance}\label{sec:Feshbach}

\begin{figure}[!b]
    \centering
    \includegraphics[width=\columnwidth]{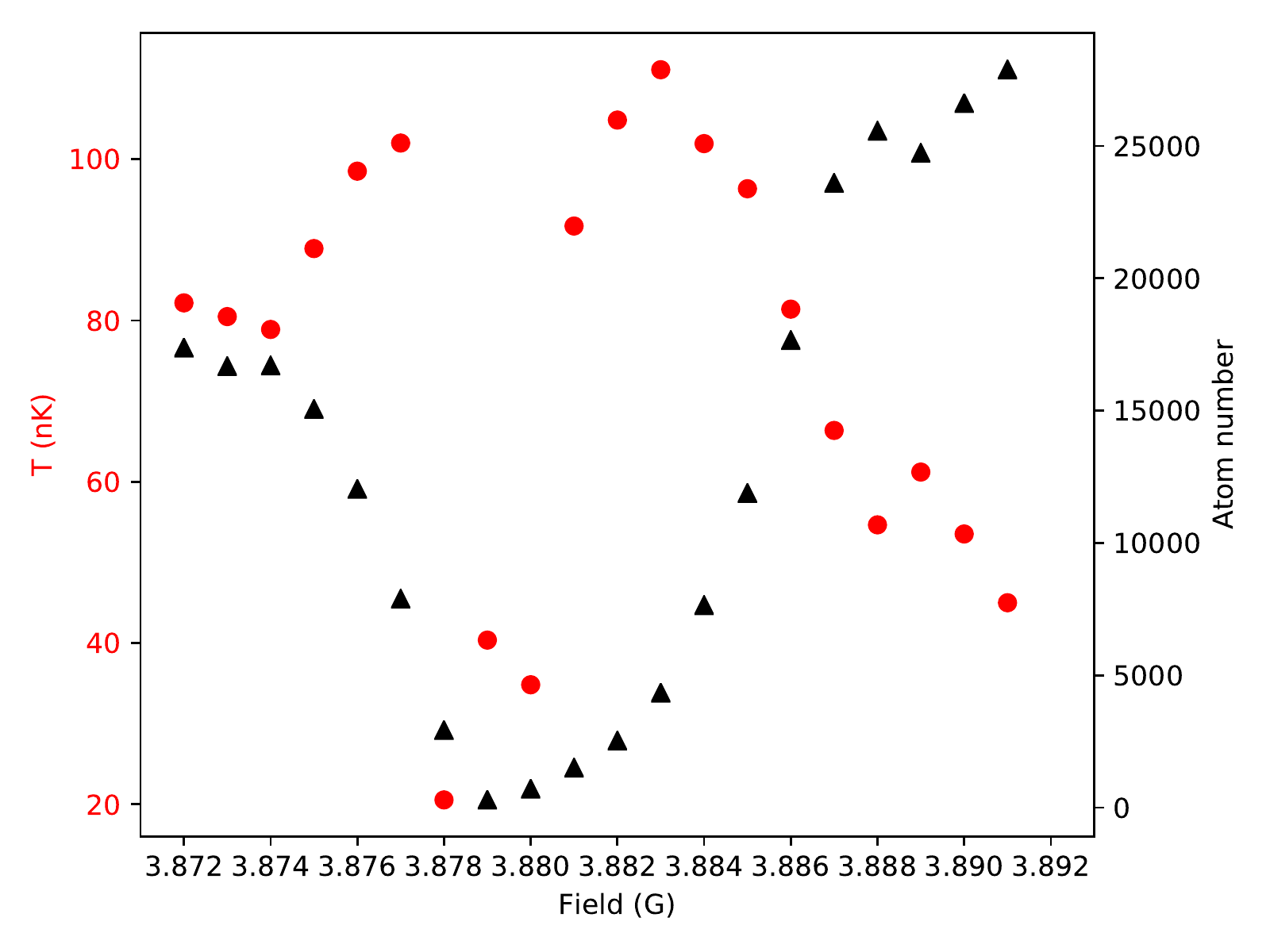}
    \caption{Atom loss and temperature spectroscopy at fields near 3.88~G. Atom number shows a minimum at $B_{02}=$~3.879(1)~G, the pole of the resonance. The temperature has a peak at $B_{02}+\Delta_2=$~3.883~G, with $\Delta_{2}=$~4(1)~mG.  The peak arises due to the vanishing cross section, which causes worse evaporative cooling efficiency and higher final temperature. The temperature dips near the resonance pole are due to atom loss. }
    \label{fig:zero_crossing}
\end{figure}

Two Feshbach resonances are used for tuning the final $\gamma$ value in the experiments. The $s$-wave scattering length $a_\text{3D}(B)$ near each Feshbach resonance can be modeled by
\begin{equation}
    a_\text{3D}(B) = a_\text{bg}\left( 1-\sum_i\frac{\Delta_i}{B-B_{0i}} \right),
\end{equation}
for resonance poles at $B_{0i}$ with widths $\Delta_i$.  The background scattering length is $a_\text{bg}$.  Table~\ref{tab:FR_parameters} lists the parameters of the two Feshbach resonances. These parameters are obtained from anisotropic expansion measurements~\cite{Tang2016aeo} and atom loss \& temperature measurements during evaporative cooling versus $B$-field, as shown in Supp.~Fig.~\ref{fig:zero_crossing}. The field at the point of greatest atom loss and the highest temperature indicate the pole and zero-scattering-length fields, respectively.  From these we infer $B_{02}$ and $\Delta_2$. For the data in Fig.~4 of the main text, BEC evaporation is performed at 4.8~G for $\gamma_\text{T}(55^\circ)=6.7$ and $16$, and at 3.95~G for $\gamma_\text{T}(55^\circ)=3.2$.  Here, $\gamma_T$ is our estimate of the average $\gamma$ in the finite-temperature experimental setup.  We discuss the calculation of $\gamma_T$ below. For the TG data, the Feshbach resonance near 27~G is used to produce the high-$\gamma$ state; see Ref.~\cite{Kao2021tpo}.
\begin{table}[t!]
    \centering
    \begin{tabular}{ccccc}
    \hline
        $a_\text{bg} (a_0)$ & $B_{01}$ (G) & $\Delta_1$ (mG) & $B_{02}$ (G) & $\Delta_2$ (mG)\\
    \hline
        156(4) & 5.112(1) & 24(2) & 3.879(1) & 4(1)
    \end{tabular}
    \caption{Feshbach resonance parameters.}
    \label{tab:FR_parameters}
\end{table}

\subsection{Time-of-flight imaging}\label{sec:TOF}

We use time-of-flight (TOF) absorption imaging. To ensure 3D ballistic expansion without contact interaction effects, we quickly change the $B$-field to set $a_\text{3D} \approx 0$ in less than a ms before release.  An 18-ms 3D TOF expansion is used for all datasets. The resulting TOF images contain momentum/rapidity information along the $\hat{x}$ direction.

\subsection{1D expansion}
Mapping rapidities to free-particle momenta requires a long 1D expansion time. To expand the gas in 1D, the 1064-nm ODT beam is suddenly turned off. After the gas expands for a time $t_\text{1D}$ in the quasi-1D geometry, we perform the same measurement as described above.

\subsection{Momentum focusing}
\begin{figure}[t!]
    \centering
    \includegraphics[width=\columnwidth]{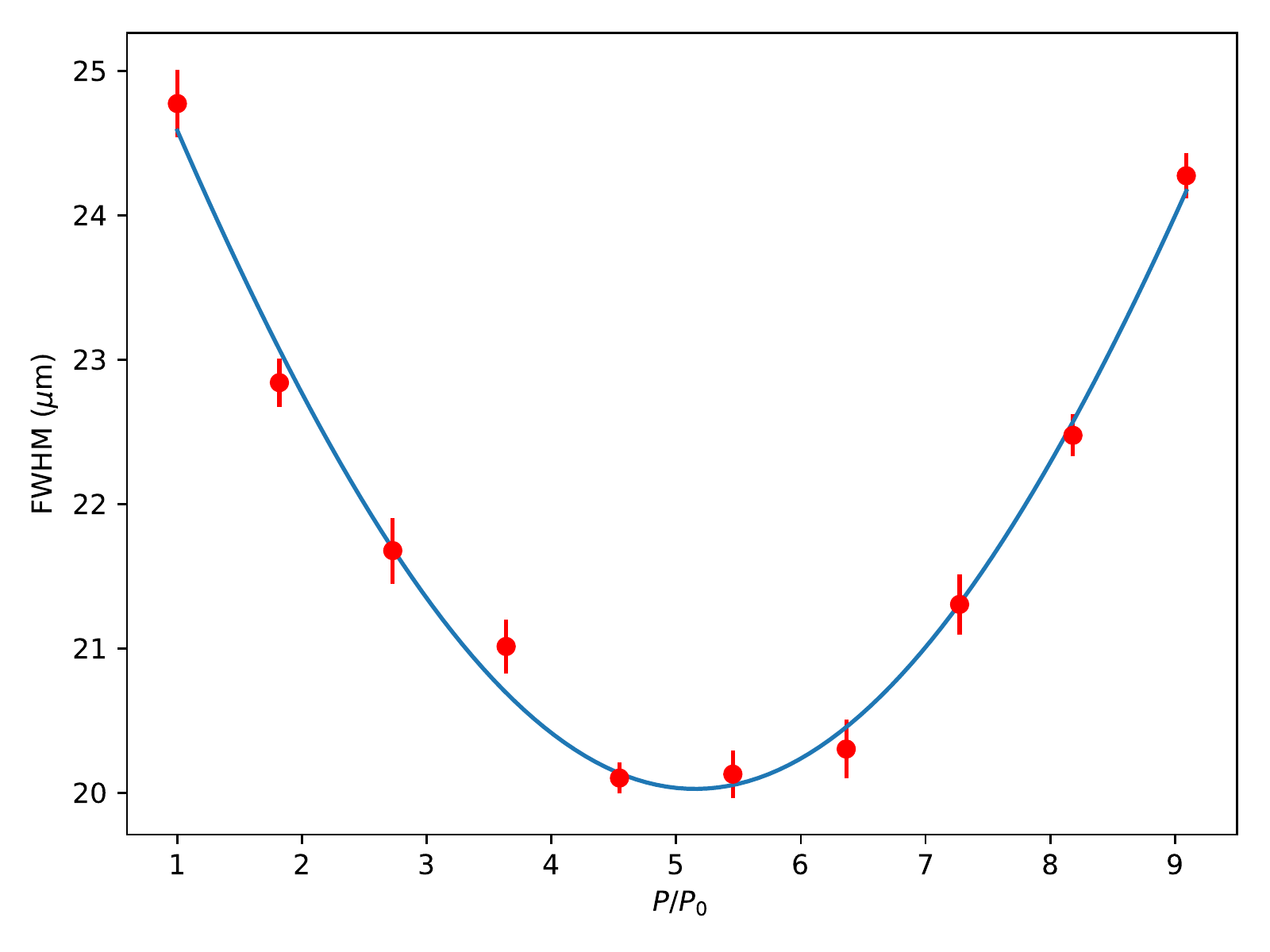}
    \caption{Measured FWHM as a function of the focusing power. The field angle is set to $\theta_\text{B}=55^\circ$. The fitted curve indicates that optimal focusing occurs near $P/P_0=5.2$.}
    \label{fig:momentum_focusing_calibration}
\end{figure}

The long initial size of the unexpanded 1D gas can distort the TOF images, resulting in a poor momentum distribution measurement. We employ a momentum focusing protocol to correct this. A short momentum kick is applied to the gas along the longitudinal direction for 0.4~ms. This occurs via a sudden jump of power in the 1064-nm ODT after $a_\text{3D}$ is set to zero. The exact power level is calibrated experimentally, as shown in Fig.~\ref{fig:momentum_focusing_calibration}. The measured full-widths at half-maximums (FWHMs) are fitted to $w=w_0\sqrt{1+\alpha (P/P_0-P_\text{focus})^2}$, where $P_0$ is the original ODT laser power; $P_\text{focus}$ is found to be approximately $5.2 P_0$.  We do not need to use momentum focusing for the rapidity measurements after 1D expansion because the longer total expansion time makes the initial size effect negligible.  This 1D expansion time is $t_\text{1D}=15$~ms; see discussion in main text.

\begin{figure}[!t]
    \centering
    \includegraphics[width=\columnwidth]{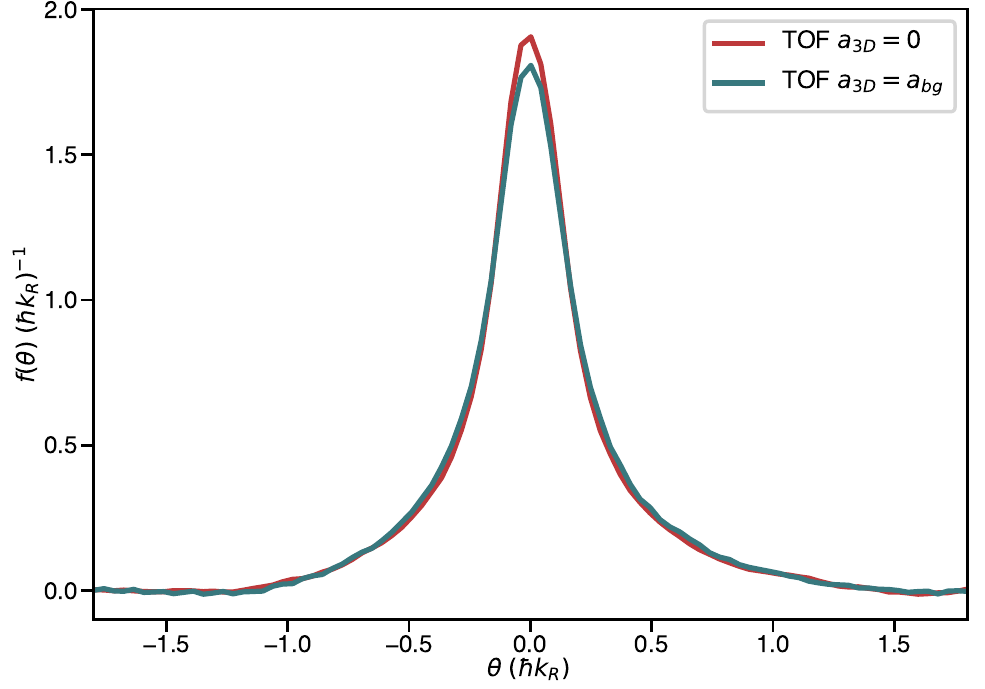}
    \caption{Red (green) curve shows the 1D distribution of the gas after momentum focusing with the $s$-wave interaction turned off (or remaining at $a_{bg}$) during the momentum kick. Here, $\theta_\text{B}=55^\circ$.}
    \label{fig:a_3D_effect}
\end{figure}

The interaction between atoms can complicate the effect of the momentum kick in the momentum focusing protocol. We experimentally show that there is a broadening effect when the $s$-wave interaction is not turned off; see Fig.~\ref{fig:a_3D_effect}. The low-momentum part near the center is broadened due to the $s$-wave interaction, showing that we must turn off this interaction before momentum focusing.

\subsection{Data analysis}

\subsubsection{Image processing}

The absorption images are processed by eigenimage analysis to remove fringes due to shot-to-shot variations of the imaging beam. All the processed images are post-selected for $\pm 10\%$ of the mean atom number. The column density then provides the 1D momentum/rapidity distributions. Between 20-to-30 shots are typically taken for each configuration setting. A pixel-wise optical density (OD) average and standard deviation ($\sigma_\text{OD}$) is calculated. We flag pixels that are part of the zero-atom region if they are within $\pm 1 \sigma_\text{OD}$ of zero. We then fit this zero-atom region with a third-order polynomial to determine the background noise level. (A small amplitude sinusoidal function is added to the fit to account for a periodic variation in background.) The background is then subtracted from the distribution. The FWHM of each 1D distribution is determined from each average. The lower (upper) bound of FWHM is estimated by offsetting the zero point of the distribution to the max (min) value of the residual background. Their differences from the nominal FWHM are calculated, and the larger value is reported as the FWHM uncertainty in the inset to Fig.~2 of the main text.

\subsection{Heating during the 1D gas preparation}

We now gauge the amount by which the system heats during the loading procedure.  To do so, we load the BEC into the 2D optical lattice up to various lattice depths and then follow the exact same reverse procedure to remove the 2D optical lattice. We then use a double-Gaussian fit to obtain a measure of the temperature from the Gaussian-shaped thermal wings. A Gaussian, rather than inverted parabola is used to fit the center, because the BEC is not deeply in the Thomas-Fermi regime. The measured temperature is corrected via deconvolution by two factors arising from the finite TOF and the finite-imaging resolution (${\sim}4~\mu$m). 

\begin{figure}[!t]
    \centering
    \includegraphics[width=\columnwidth]{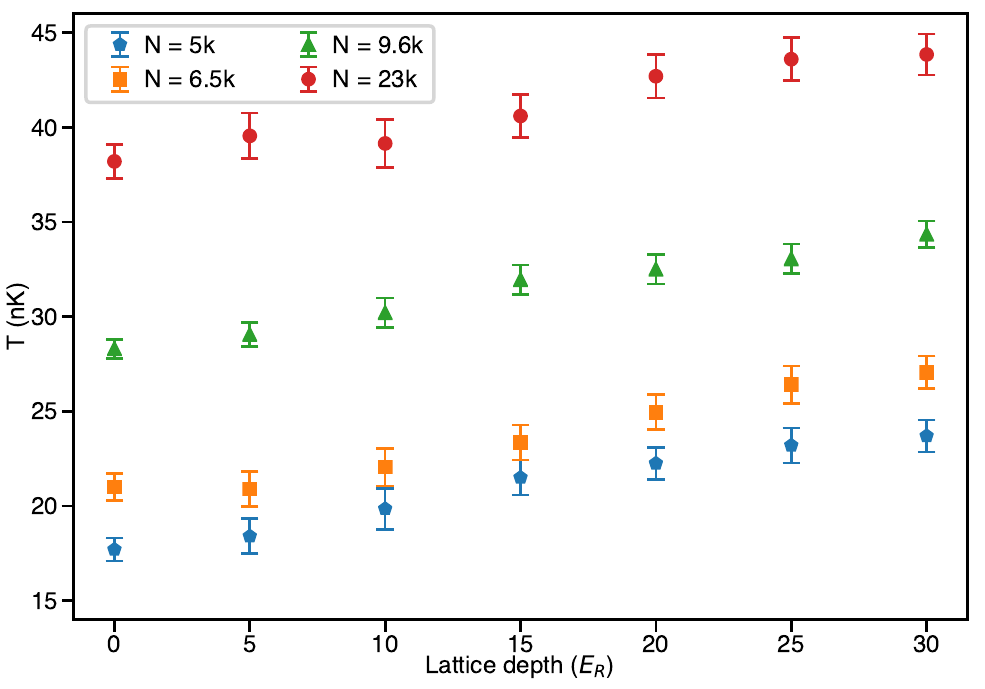}
    \caption{Inferred temperature versus lattice depth during the loading sequence. Red circles represent the data for the number of atoms used for Fig.~4 in the main text. Orange squares and blue pentagons show the measurements with atom numbers slightly above and below that which we use in the TG-gas experiments.}
    \label{fig:loading_temperature}
\end{figure}

It is assumed that the temperature increase is identical during loading and deloading, allowing us to infer the amount of heating that occurs during the loading procedure. The results are presented in Fig.~\ref{fig:loading_temperature} for various initial atom numbers. Because the results show only a few nK of heating, we neglect the effect of heating after the 1D gases decouple; see the modeling discussion below. If the system is always thermalizing, any potential heating that occurs before the decoupling of the system into 1D gases is expected to be accounted for by our fitting of the temperature of the 1D gases when they decouple (see the modeling discussion next).

\section{Modeling}

\subsection{Experimental system}

Neglecting the intertube DDI, the experimental system can be modeled as a 2D array of independent 1D gases. Each 1D gas is described by the Lieb-Liniger Hamiltonian~\cite{lieb_liniger_63}, with the addition of an intratube DDI $U^{\rm 1D}_{\rm DDI}$ and a longitudinal harmonic confining potential $U_{\rm H}$,
\begin{align}\label{eq:H_LL1}
H&=\sum_{i=1}^{N}\left[-\frac{\hbar^2}{2m}\frac{\partial^2}{\partial x_i^2}+U_{\rm H}(x_i)\right]\nonumber\\&+\sum_{1\leq i<j \leq N}\left[g_{\rm 1D}^{\rm vdW}\delta(x_i-x_j)+U^{\rm 1D}_{\rm DDI}(\theta_\text{B}, x_i-x_j)\right]\,,
\end{align}
where $m$ is the mass of the $^{162}$Dy atoms and $N$ is the number of atoms. $g_{\rm 1D}^{\rm vdW}=-\hbar^2/(ma_{\rm 1D})$ is the effective 1D contact interaction due to the van der Waals force. It depends on the depth of the 2D optical lattice $U_{\rm 2D}$ and on the 3D s-wave scattering length $a_{\rm 3D}(B)$, set by the magnetic field, through $a_{\rm 1D}=-a_{\perp}(a_{\perp}/a_{\rm 3D}(B)-C)/2$.  $a_{\rm 1D}$ is the 1D scattering length, where $a_{\perp}=\sqrt{2\hbar/(m\omega_{\perp})}$ is the transverse confinement, $\omega_{\perp}=\sqrt{2U_{\rm 2D}k_0^2/m}$, and $C=-\zeta(1/2)\simeq1.4603$~\cite{olshanii_98}. The longitudinal confinement potential is modeled as a harmonic trap $U_{\rm H}(x)=\frac{1}{2}m\omega_x^2 x^2$, where $\omega_x$ is the trapping frequency.

The effective intratube DDI in the single-mode approximation has the following form~\cite{Deuretzbacher2010gpo,Deuretzbacher2013egp,Tang2018tni,Kao2021tpo,DePalo2021pad},
\begin{equation}
U^{\rm 1D}_{\rm DDI}(\theta_\text{B}, x)=\frac{\mu_0\mu^2}{4\pi}\frac{1-3\cos^2\theta_\text{B}}{\sqrt{2}a_{\perp}^3}\bigg[V^{\rm 1D}_{\rm DDI}(u)-\frac{8}{3}\delta(u)\bigg]\,,
\end{equation}
where $V^{\rm 1D}_{\rm DDI}(u)=-2|u|+\sqrt{2\pi}(1+u^2)e^{u^2/2}{\rm erfc}(|u|/\sqrt{2})$, $u=\sqrt{2}x/a_{\perp}$, and ${\rm erfc}(x)$ is the complementary error function. $\mu=9.93\mu_\text{B}$ is the dipole moment of $^{162}$Dy. 

The 1D Hamiltonian~\eqref{eq:H_LL1} is not exactly solvable in the presence of $U^{\rm 1D}_{\rm DDI}(\theta_\text{B}, x)$, except at $\theta_\text{B}\approx55^\circ$, at which the intratube DDI vanishes. In order to account for the leading-order effect of the intratube DDI, which is due to its short-range part, we consider it as a correction to the contact interaction term, 
\begin{equation}\label{UthetaB}
\tilde U(\theta_\text{B}, x)=\frac{\mu_0\mu^2}{4\pi}\frac{1-3\cos^2\theta_\text{B}}{2a_{\perp}^2}\left[A-\frac{8}{3}\right]\delta(x)=g_{\rm 1D}^{\rm DDI}\delta(x)\,,
\end{equation} 
where $A=\int_{-\infty}^{\infty}V^{\rm 1D}_{\rm DDI}(u)du=4$. The Hamiltonian~\eqref{eq:H_LL1} can then be written as
\begin{equation}\label{eq:H_LL2}
\tilde H=\sum_{i=1}^{N}\bigg[-\frac{\hbar^2}{2m}\frac{\partial^2}{\partial x_i^2}+U_{\rm H}(x_i)\bigg]+\sum_{1\leq i<j \leq N}g_{\rm 1D}\delta(x_i-x_j)\,,
\end{equation}
where 
\begin{equation}\label{eq:g1d}
g_{\rm 1D}=g_{\rm 1D}^{\rm vdW}+g_{\rm 1D}^{\rm DDI}.
\end{equation}
The values of $g_{\rm 1D}^{\rm vdW}$, $g_{\rm 1D}^{\rm DDI}$, and $g_{\rm 1D}$ for the experimental results reported in the main text are shown in Table~\ref{table:g}. In the absence of $U_{\rm H}$, this Hamiltonian is exactly solvable via the Bethe ansatz~\cite{lieb_liniger_63,yang_yang_69}. The observables in equilibrium depend on the dimensionless parameter $\gamma=mg_{\rm 1D}/(\hbar^2n_{\rm 1D})$, where $n_{\rm 1D}$ is the 1D density.

\begin{table}
\centering
\begin{tabular}{cccccc}
$a_{\rm 3D}$  & $g_{\rm 1D}^{\rm vdW}$  & $\theta_\text{B}$ & $g_{\rm 1D}^{\rm DDI}$  & $g_{\rm 1D}$ \\
($a_0$) & ($\frac{\hbar^2}{m}$ $\mu$m$^{-1}$) & & ($\frac{\hbar^2}{m}$ $\mu$m$^{-1}$) & ($\frac{\hbar^2}{m}$ $\mu$m$^{-1}$) \\[1ex] \hline
89.5     & 4.1      & $0^\circ$  & -5.4     & -1.3 \\
89.5     & 4.1      & $35^\circ$ & -2.7     & 1.4  \\
89.5     & 4.1      & $55^\circ$ & 0        & 4.1  \\
89.5     & 4.1      & $90^\circ$ & 2.7      & 6.8  \\
167      & 8.5      & $0^\circ$  & -5.4     & 3.1  \\
167      & 8.5      & $35^\circ$ & -2.7     & 5.8  \\
167      & 8.5      & $55^\circ$ & 0        & 8.5  \\
167      & 8.5      & $90^\circ$ & 2.7      & 11.2 \\
320      & 20.4     & $0^\circ$  & -5.4     & 15.0 \\
320      & 20.4     & $35^\circ$ & -2.7     & 17.7 \\
320      & 20.4     & $55^\circ$ & 0        & 20.4 \\
320      & 20.4     & $90^\circ$ & 2.7      & 23.1 \\
803      & 260      & $90^\circ$ & 2.7      & 263     
\end{tabular}
\caption{List of $g_{\rm 1D}$ parameters. Relevant scattering lengths and the corresponding $g_{\rm 1D}^{\rm vdW}$ are listed. The DDI correction $g_{\rm 1D}^{\rm DDI}$ is calculated using Eq.~\eqref{UthetaB}, and $g_{\rm 1D}$ is calculated using Eq.~\eqref{eq:g1d}.}\label{table:g}
\end{table}

We use path integral quantum Monte Carlo with worm updates~\cite{boninsegni_prokofev_06a, boninsegni_prokofev_06b, xu_rigol_15} to compute the momentum distributions for the equilibrium states of Hamiltonian~\eqref{eq:H_LL2}. The rapidity distributions for these states are calculated solving the thermodynamic Bethe ansatz equations~\cite{yang_yang_69} within the local density approximation (LDA).

\section{Simulation of state preparation}\label{sec:statprep}

To compare to the experimental measurements, we carry out theoretical calculations that model the 1D gases forming the 2D array. For 1D gas ``$l$,'' at spatial location $(y_l, z_l)$ in the 2D plane, we need to find the number of atoms $N_l$ and the temperature $T_l$. In this section, we explain how these parameters are computed to model as close as possible what is experimentally done.

In our experiment, we load a 3D BEC with $N_{\rm tot}$ atoms at temperature $T_{\rm 3D}$ in a 2D optical lattice ($U_{\rm 2D}$) that is ramped-up to create the 2D array of 1D gases. Our {\it first} theoretical assumption is that at a lattice depth $U^*_{\rm 2D}$ (which sets $g^*_{\rm 1D}$) the 3D gas decouples into individual 1D gases with $N_l$ atoms. Our {\it second} assumption is that all the 1D gases are in thermal equilibrium at a temperature $T^*$ with each other at that point. Given those assumptions, we can use the exact solution of the homogeneous Lieb-Liniger model at finite temperature---together with the LDA and knowledge from measurements of $N_{\rm tot}$ and ODT frequencies $\omega_x$, $\omega_y$, and $\omega_z$---to determine the number of atoms $N_l$ in each 1D gas.  (We round $N_l$ to the nearest integer number.) We can do this for any given decoupling depth $U^*_{\rm 2D}$ and temperature $T^*$ at decoupling. We also compute the entropy $S_l$ of each 1D gas at that point. The effectiveness of such a theoretical modeling procedure, under the assumption that the entire system is in the ground state at all times, was demonstrated in Ref.~\cite{malvania_zhang_21} for 1D gases with only contact interactions.

As the loading in the 2D optical lattice proceeds beyond $U^*_{\rm 2D}$ to the final 2D lattice depth, our {\it third} assumption is that the process is adiabatic; namely, that the entropy $S_l$ (not the temperature) of each 1D gas remains constant. Using that entropy and the parameters for each 1D gas at the end of the state preparation, we find the temperature $T_l$ of each 1D gas and use it to compute the momentum and rapidity distributions. It is important to stress that after the 2D lattice is ramped to its final depth, our state preparation includes the experimentally relevant reduction of the longitudinal trapping frequency $\omega_x$. This overall procedure not only makes $T_l$ lower than $T^*$ but also results in different temperatures across the array of 1D gases. In our calculations, we first group the 1D tubes with the same $N_l=N$ (rounded to the closest integer) and assume that they have the same entropy [using the average entropy $\bar S(N) = \overline{ S_l(N_l=N) }$]. Then, we search for $T(N)$ (in a grid of temperatures that change in steps of 0.5~nK) that produces the appropriate $\bar S(N)$ using the value of $\omega_x$ and $g_{\rm 1D}$ at the end of the state preparation.

The free parameters in our modeling of the experiment are $U^*_{\rm 2D}$ and $T^*$. We select their values by minimizing the quadrature sum of the differences between the experimental measurements and the theoretical calculations for the momentum and rapidity distributions
\begin{equation}
\Delta=\sqrt{\Delta^2_{\rm rapidity}+\Delta^2_{\rm momentum}}\,,
\end{equation}
where 
\begin{equation}
\Delta_{\alpha}=\frac{\sum |f_{\alpha}^{\rm exp.}(k)-f_{\alpha}^{\rm theo.}(k)|\delta k}{\sum |f_{\alpha}^{\rm exp.}(k)|\delta k+\sum |f_{\alpha}^{\rm theo.}(k)|\delta k}\,,
\end{equation}
and $\alpha$ denotes either rapidity or momentum. We call $\Delta$ the theoretical error.

\begin{figure}[!t]
    \centering
    \includegraphics[width=\columnwidth]{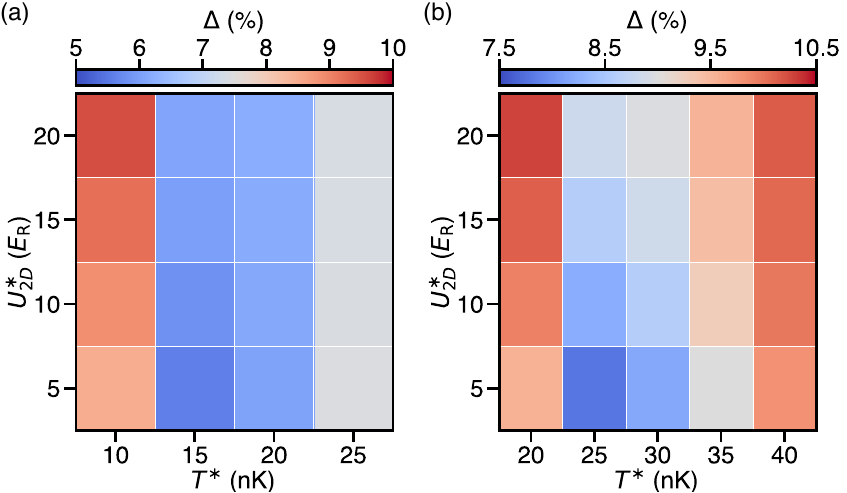}
    \caption{Theoretical error for the experimental parameters considered in Fig.~2 of the main text. (a) and (b) correspond to the parameters in the same panels of that figure.}
    \label{fig:parameter optimization}
\end{figure}

Results for the theoretical error as a function of $T^*$ and $U^*_{\rm 2D}$ are shown in Fig.~\ref{fig:parameter optimization}. We use a coarse grid of temperatures (in 5-nK steps) and decoupling depths (in $5E_\text{R}$ steps) because our strong assumptions in the modeling of the state preparation do not make a finer grid quite meaningful. Figure~\ref{fig:parameter optimization} shows that, around the minimum theoretical error, the error depends strongly on $T^*$ and weakly on $U^*_{\rm 2D}$. The latter suggests that assuming a single decoupling depth is a reasonable approximation. We also find that, likely because of ignoring the effect of the DDI, our theoretical error is lower at low decoupling depths. This is because low decoupling depths result in higher particle densities in the 1D gases and, hence, in broader rapidity distributions, which are closer to (yet still narrower than) the experimental ones. For our simulations we select $U^*_{\rm 2D}=5E_\text{R}$. The line cut along $5E_\text{R}$ is shown in the insets of Fig.~3 of the main text.

An interesting observation is that the experimentally estimated BEC temperature $T_{\rm 3D}$ is higher than the optimal $T^*$ at decoupling, which suggests that the loading from 3D to 1D results in cooling even if this process is not perfectly adiabatic---we do not assume adiabaticity in the transition from 3D to 1D.  Note that adiabaticity is assumed after decoupling.

For the results reported in the main text, the intertube DDI energy is estimated by summing all of the DDI energy between atoms in the 2D array of 1D gases using the density profiles calculated from the state preparation model.

\subsection{Reduction of $\omega_x$}\label{sec:lowerw}

Our modeling of the experimental state preparation pinpointed a process that results in cooling of the 1D gases (a reduction in $T_l$ from $T^*$). It is the ramping down of the longitudinal trap frequency with the opening of the trap; $\omega_x$ changes linearly from $\omega_x^{i}=2\pi\times$55.5~Hz to $\omega_x^{f}=2\pi\times$36.4~Hz in 150~ms. The 1D gases expand and adiabatically cool. 

\begin{figure}[!t]
\includegraphics[width=0.95\columnwidth]{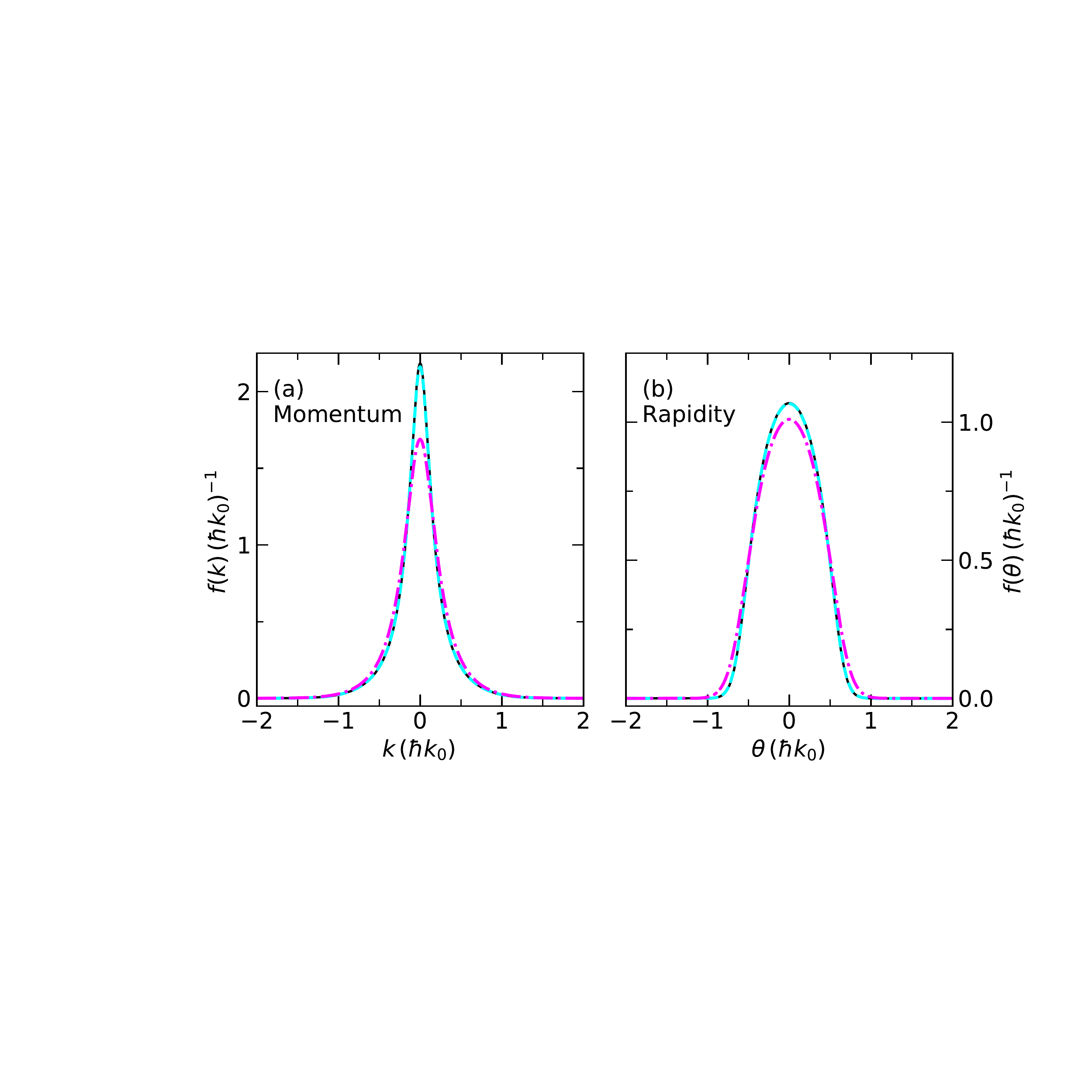}
\caption{Numerical results for (a) momentum and (b) rapidity distributions after the reduction of $\omega_x$ to its final value. We studied a single 1D gas with $N=20$ particles at the TG limit. Solid lines are the results of the exact dynamics following the reduction of $\omega_x$ in the experiment, where we start from an equilibrium state at 15~nK with $\omega_x^{i}=2\pi\times55.5$~Hz and simulate the linear reduction of the trap frequency to $\omega_x^{f}=2\pi\times36.4$~Hz in 150~ms. Dashed (dashed-dotted) lines are the calculations for an equilibrium state at 10~nK (15~nK) with $\omega_x=2\pi\times36.4$~Hz. The 10~nK state is the one that has the closest entropy (within the 0.5-nK temperature-resolution used) to the initial state.}
\label{fig:TG_trapturndown}
\end{figure}

We used numerical simulations in the TG limit ($g_{\rm 1D}\to\infty$) to check that the experimental protocol followed during this step is slow enough to produce adiabatic cooling. In Fig.~\ref{fig:TG_trapturndown}, we plot the numerical results for the (a) momentum and (b) rapidity distributions after the reduction of $\omega_x$. The calculations were done for a single 1D gas with $N=20$ atoms at $T^{i}=15$~nK before the trap changes. We do the exact time evolution with the numerical approach introduced in Ref.~\cite{Xu2017eoo}. The linear ramp downward of the trap frequency is discretized in 600 steps in our calculation. The results after the linear ramp (solid lines) agree with the calculations for an equilibrium state in the $\omega_x=2\pi\times36.4$~Hz trap at 10~nK (dashed lines), which has the closest entropy (within the 0.5-nK temperature-resolution used) to the state ($T^{i}=15$~nK with $\omega_x=2\pi\times55.5$~Hz) before this operation. As reference, we also plot calculations for an equilibrium state in the $\omega_x=2\pi\times36.4$~Hz trap at 15~nK (dashed-dotted lines). \\

\pagebreak
\section{Definition and values of $\gamma_T$}

\begin{table}[b!]
\begin{tabularx}{0.45\textwidth}
{  >{\centering\arraybackslash}X 
   >{\centering\arraybackslash}X 
   >{\centering\arraybackslash}X 
   >{\centering\arraybackslash}X 
   >{\centering\arraybackslash}X  }
 $g_{\rm 1D}$& $\bar T$ &$\gamma_{\rm T}$ \\
   ($\hbar^2/m\,\mu$m$^{-1}$) & (nK) &  \\
 \hline
 -1.3 & -  & -\\
 1.4 & 16 & 1.0 \\
 4.1 & 16 & 3.2\\
 6.8 & 16 & 5.3\\
 3.1 & 16  &2.2\\
 5.8 & 16 & 4.5\\
 8.5 & 16.5 & 6.7 \\
 11.2 & 16.5 & 8.7\\
 15.0 & 16.5 & 12\\
 17.7 & 16.5 & 14\\
 20.4 & 17 & 16\\
 23.1 & 17 & 19\\
 263 & 10 & 420
\end{tabularx}
\caption{\label{tab:gammaT} Calculated $\gamma_\text{T}$ for the experimental parameters considered in the main text. See Table~\ref{table:g} for the experimental parameters underpinning the calculation of $g_\text{1D}$.}
\end{table}

Due to the presence of the confining potential in experimental systems, $\gamma(x)=mg_{\rm 1D}/[\hbar^2n_{\rm 1D}(x)]$ depends on the local 1D density $n_{\rm 1D}(x)$. To gauge how strongly correlated a specific inhomogeneous 1D system is, one can compute the  weighted average $\bar \gamma = \int dx n_{\rm 1D}(x) \gamma(x) / [\int dx n_{\rm 1D}(x)] = {\int dx g_{\rm 1D}}/N$.  $\bar\gamma$ is well defined at zero temperature, $\int dx=x_0$, where $x_0$ is the size of the trapped 1D gas; however, at finite temperature, the particle density exhibits long tails and so $x_0$ is not well defined~\cite{xu_rigol_15}. Previous work circumvented this problem replacing $x_0\rightarrow x^\eta_0$, where $ x^\eta_0$ is computed using the specific fraction $\eta$ of the atoms that is within $[-x^\eta_0,x^\eta_0]$; e.g., 80\% of the atoms in Ref.~\cite{Jacqmin2012mdo}. We follow a different approach in order to avoid the arbitrariness in selecting the fraction $\eta$.

We instead report $\gamma_T$, which is based on the ratio of kinetic and interaction energy. We compute $\gamma_T$ as follows. For a given set of experimental parameters, we calculate the ratio between the total kinetic ($E_K=\sum_l E^l_K$) and total interaction energy ($E_I=\sum_l E^l_I$), as obtained from our modeling. $\gamma_T$ is the value of $\gamma$ of a homogeneous system at finite temperature that has exactly that ratio. The homogeneous system is selected to have the same $g_{\rm 1D}$ as the trapped one, and a temperature that is the weighted average temperature of the array of 1D gases, $\bar T=\sum_l N_l T_l/\sum_l N_l$. Since $E_K/E_I$ is a monotonic function with $\gamma$, it is straightforward to find the particle density $n^{\bar T}_{\rm 1D}$ in such a homogeneous system for which $E_K/E_I$ matches the result obtained for the modeling of the experimental results. Then we compute $\gamma_T=mg_{\rm 1D}/(\hbar^2n^{\bar T}_{\rm 1D})$. In Table \ref{tab:gammaT}, we show the calculated $\gamma_T$ for the experimental parameters considered in the main text.

\end{document}